\documentclass[twocolumn,showpacs,preprintnumbers,superscriptaddress,amsmath,amssymb]{revtex4-1}
\usepackage[dvips]{graphicx,color}
\usepackage{dcolumn}
\usepackage{bm}
\usepackage{ulem}

\begin{document}
\def\bea{\begin{eqnarray}}
\def\eea{\end{eqnarray}}
\def\be{\begin{equation}}
\def\ee{\end{equation}}
\def\rra{\right\rangle}
\def\lla{\left\langle}
\def\rv{\bm{r}}
\def\la{\Lambda}
\def\sgm{\Sigma^-}
\def\eps{\epsilon}
\def\ms{M_\odot}
\def\beff{B_{\rm eff}(\rho_B)}
\def\qc{\rho_{\rm ch}}
\def\fv{f_V}

\preprint{APT/123-QED
\hspace{0.2cm}
ZTF-EP-14-04}

\title{ 
Finite-size effects at the hadron-quark transition and heavy hybrid stars 
} 

\author{Nobutoshi Yasutake}
 \email{nobutoshi.yasutake@it-chiba.ac.jp}
 \affiliation{Department of Physics, Chiba Institute of Technology, 
 2-1-1 Shibazono,Narashino, Chiba 275-0023, Japan}
\author{Rafa{\l} {\L}astowiecki}
 \email{lastowiecki@ift.uni.wroc.pl}
 \affiliation{Institute for Theoretical Physics, University of Wroc{\l}aw, 
 Max Born pl. 9, 50-204 Wroc{\l}aw, Poland}
\author{Sanjin Beni{\'c}}
 \email{sanjinb@phy.hr}
 \affiliation{Physics Department, Faculty of Science, University of Zagreb, 
 Zagreb 10000, Croatia}
\author{David Blaschke}
 \email{blaschke@ift.uni.wroc.pl}
 \affiliation{Institute for Theoretical Physics, University of Wroc{\l}aw, 
 Max Born pl. 9, 50-204 Wroc{\l}aw, Poland}
 \affiliation{Bogoliubov Laboratory for Theoretical Physics, JINR Dubna,
 141980 Dubna, Russia}
\author{Toshiki Maruyama}
 \email{maruyama.toshiki@jaea.go.jp}
 \affiliation{Advanced Science Research Center, Japan Atomic Energy Agency, 
 Tokai, Ibaraki 319-1195, Japan}
\author{Toshitaka Tatsumi}
 \email{tatsumi@ruby.scphys.kyoto-u.ac.jp}
\affiliation{Department of Physics, Kyoto University, Kyoto 606-8502, Japan}

\date{\today}

\begin{abstract}
We study the role of finite-size effects at the hadron-quark phase transition 
in a new hybrid equation of state constructed from an ab-initio 
Br\"uckner-Hartree-Fock equation of state with the realistic Bonn-B potential 
for the hadronic phase and a covariant non-local Nambu--Jona-Lasinio model
for the quark phase. 
We construct static hybrid star sequences and find that our model can support 
stable hybrid stars with an onset of quark matter below $2~M_\odot$ and a 
maximum mass above $2.17~M_\odot$ in agreement with recent observations.
If the finite-size effects are taken into account the core is composed
of pure quark matter.
Provided that the quark vector channel interaction is small, and the finite 
size effects are taken into account, quark matter appears at densities 2-3 
times the nuclear saturation density.
In that case the proton fraction in the hadronic phase remains below the value 
required by the onset of the direct URCA process, so that the early onset
of quark matter shall affect on the rapid cooling of the star.
\end{abstract}

\maketitle

\section{\label{sec:level1}Introduction}
The equation of state~(EoS) is
the central quantity for the study of compact stars. 
Since modern lattice QCD simulations are not applicable at large baryon 
densities and low temperatures $T\simeq 0$, there is a large uncertainty in 
theoretical descriptions of the behavior of matter at extreme densities.  
The understanding may be improved by studying astrophysical phenomena; 
namely, we may use the known astrophysical constraints from observations of 
compact stars in order to provide constraints on the EoS. 
Recently, the idea has been pursued to use a Bayesian analysis (BA) for 
``inversion'' of the Tolman-Oppenheimer-Volkoff equations, 
i.e. to extract a probability measure for models of the cold EoS in the 
pressure-energy density plane from observational data related to masses and 
radii of compact stars. 
While first analyses of this type have favored burst sources with rather 
uncertain and model dependent statements about radii 
\cite{Steiner:2010fz,Steiner:2012xt}, a very recent BA uses a set of 
stronger and statistically independent observations, testing also the 
possibility of a first order phase transition at supersaturation densities 
\cite{Blaschke:2014via}.

At this point the strongest restriction to the EoS is provided by the recent 
measurement of the high mass of $\sim  2~M_\odot$ from observations of the 
pulsars PSR J1614-2230 by Demorest et.~al.~\cite{Demorest:2010bx} and 
PSR J0348+0432 by Antoniadis et.~al.~\cite{Antoniadis:2013pzd}.
The recent BA \cite{Blaschke:2014via} makes use of this constraint together 
with a new mass-radius constraint from the precise timing analysis of the 
nearest known millisecond pulsar PSR J0437-4715 \cite{Bogdanov:2012md} and 
the constraint on the gravitational binding for the neutron star B in the 
binary system J0737-3039(B) \cite{Podsiadlowski:2005ig}, see also 
\cite{Kitaura:2006}, at the precisely measured gravitational mass of 
$1.249 \pm 0.001~M_\odot$.

There are many studies relating astrophysical phenomena involving compact stars
and the properties of matter at extreme densities, eventually including the 
possibility of a quark deconfinement transition. 
These concern, e.~g., the cooling of compact 
stars~\cite{Page:2000wt,Blaschke:1999qx,Blaschke:2000dy,Grigorian:2004jq}, 
gravitational wave emission
\cite{Lin:2005zda,Yasutake:2007st,Abdikamalov:2008df}, 
neutrino emission~\cite{Hatsuda:1987ck,Nakazato:2008su,Sagert:2008ka}, 
eigenfrequencies~\cite{Burgio:2003mr}, 
and the energy release during the collapse of neutron stars to quark 
stars~\cite{Aguilera:2002dh,Yasutake:2004kx,Zdunik:2006uw}.

The study of the baryon-baryon (BB) interaction in lattice QCD simulations 
recently became a hot topic ~\cite{Ishii:2006ec,Walker-Loud:2014iea}. 
Experiments like JPARC will also provide valuable information on the BB 
interaction. 
In the near future the EoS in the hadronic phase may be determined by
incorporating this information on the BB interaction in the
Br\"{u}ckner-Hartree-Fock~(BHF) theory~\cite{bhf}, 
the variational approach~\cite{Akmal:1998cf,Takano:2010zz}, or the 
Dirac-Br\"{u}ckner-Hartree-Fock~(DBHF) 
theory~\cite{Brockmann:1990cn,vanDalen:2004pn}. 
In this paper we adopt the BHF theory for hadronic matter.

Out of a several of possible models for quark matter we use the two flavor 
covariant non-local Nambu--Jona-Lasinio (nlNJL) model 
\cite{Contrera:2010kz,Benic:2013eqa} with 
vector interactions \cite{Alvarez-Castillo:2013spa}.
The advantage over the usual local version of the NJL model is due to the 
introduction of the additional gradient self-energy channel and due to the 
explicit momentum dependence of all the dressing functions of the quark 
propagator. 
Both of these improvements are well founded on lattice QCD data 
\cite{Parappilly:2005ei,Kamleh:2007ud} and Dyson-Schwinger equation studies
\cite{Bhagwat:2003vw, Fischer:2009gk, Roberts:2012sv}, and make the non-local 
NJL model a well-calibrated, effective low-energy QCD approach to the 
thermodynamics of quark matter.

The main purpose of this work is to examine the features and the astrophysical 
consequences of the mixed phase between the pure quark and hadron matter phases
by considering {\it finite-size effects}.
Taking into account the surface tension and the charge screening we 
find the non-uniform, so-called ``pasta" structures at the hadron-quark 
interface.
In this work we investigate more in detail the occurrence of pasta structures 
for the values of the surface tension $\sigma=10$ MeV fm$^{-2}$ and 
%for $\sigma=
$40$ MeV fm$^{-2}$.
For weak surface tension the EoS of the mixed phase becomes similar to the one 
of a bulk Gibbs construction, while for strong surface tension it approaches 
the result of a Maxwell construction
\cite{Voskresensky:2001jq,Voskresensky:2002hu,Endo:2005zt,Maruyama:2007ey}, 
in which the maximum
masses with the phase transition are around $1.5 M_\odot$ and
a simple bag model was used for modeling the quark phase.
This model gives a quite simple description of quark matter,
and it should be replaced by a more sophisticated one to study
more realistically the quark-hadron phase transition.
This is the aim of the present work.

We construct the hybrid EoS and the corresponding hybrid star sequences.
For the calculation of the quark matter EoS we use the following 
values for the ratio of the vector and the scalar channel couplings 
$\eta_V = G_V/G_S = 0.10$ and $\eta_V=0.20$.
Stable hybrid stars respecting the $2~M_\odot$ constraint are found in the 
case of $\eta_V=0.10$.
The bulk Gibbs construction for this case supports only a mixed phase in the 
core. 
However, taking into account finite-size effects the cores of massive hybrid stars
are composed of pure quark matter.
%However, taking into account finite-size effects 
%the core of hybrid stars is composed of pure quark matter.
For $\eta_V=0.20$ the $2M_\odot$ stars are mainly composed of hadron matter.
With the appearance of quark matter at higher densities 
the star becomes unstable.
For $\eta_V=0.10$ quark matter appears at low densities causing a reduction of 
the proton fraction at the onset of the mixed phase below the threshold value 
of $1/9$ for the onset of the direct URCA (dURCA) process in the $n-p-e$ phase,
while for $\eta_V=0.20$ the proton fraction exceeds this value.

This paper is organized as follows. 
In Sec.~\ref{msec:eos}, we outline our framework for obtaining the hybrid
EoS with pasta phase. 
Sec.~\ref{sec:results} contains numerical results for the EoS with the 
different quark-hadron mixed phase constructions including the pasta phase
as well as for the corresponding compact star sequences. 
Sec.~\ref{sec:conclude} is devoted to the conclusion and a discussion of 
some astrophysical implications of our results.

\section{Equation of state}
\label{msec:eos}
\subsection{\label{sec:mit} Equation of state for quark phase \\
---non-local NJL model}

The current theoretical description of quark matter 
includes many uncertainties, seriously limiting the predictability of the EoS 
at high baryon density. 
We resort here to a field theoretical model for the quark matter EoS 
and apply constraints on parameters from available experimental information 
and lattice QCD data.
We will use the $N_f=2$ covariant non-local Nambu-Jona--Lasinio (nlNJL) 
model \cite{Contrera:2010kz,Benic:2013eqa}.
For some of the previous works on the cold, dense 
EoS in this class of models see Ref.~\cite{Blaschke:2007ri} for superconductivity,
\cite{Orsaria:2012je} for application to 
$2+1$ flavors and \cite{Alvarez-Castillo:2013spa} where a crossover
transition was discussed at $T=0$.
  
At $T=0$ the Euclidean action is given as \cite{Contrera:2010kz,Benic:2013eqa}
\begin{eqnarray}
S_E &=& \int d^4 x  [ \bar{q}(-i\partial_\mu\gamma_\mu+m)q-i\mu\bar{q}\gamma_4 q \nonumber \\
&-&\frac{G_S}{2} \left\{ j^S_a(x)j^S_a(x)+ j_\mathbf{p}(x)j_\mathbf{p}(x)+j_{p_4}(x)j_{p_4}(x) \right\}  \nonumber \\
&+&\frac{G_V}{2}j^V_\mu(x)j^V_\mu(x) ],
\label{eq:nnjl}
\end{eqnarray}
with currents
\be
j^S_a(x)=\int d^4 z g(z)\bar{q}\left(x+\frac{z}{2}\right)
\Gamma_aq\left(x-\frac{z}{2}\right)~,
\label{eq:crt1}
\ee
\be
j_{\mathbf{p}}(x)=\int d^4 z f(z)\bar{q}\left(x+\frac{z}{2}\right)
\frac{i\overleftrightarrow{\nabla}\cdot\boldsymbol{\gamma}}
{2\kappa_\mathbf{p}}
q\left(x-\frac{z}{2}\right)~,
\label{eq:crt2}
\ee
\be
j_{p_4}(x)=\int d^4 z f(z)\bar{q}\left(x+\frac{z}{2}\right)
\frac{i\overleftrightarrow{\partial_4}\gamma_4}{2\kappa_{p_4}}
q\left(x-\frac{z}{2}\right)~,
\label{eq:crt3}
\ee
where $\Gamma_a = (1,i\gamma_5\boldsymbol{\tau})$, $\boldsymbol{\tau}$ are
Pauli matrices and $m$, $\mu$ are the current quark mass set as $m = 2.37$~MeV for u, d-quarks, and the quark chemical potential.
The vector current
\be
j^V_\mu(x) = \bar{q}(x)\gamma_\mu q(x)~,
\label{eq:veccurr}
\ee
is kept in a local form.
By different weights $\kappa_\mathbf{p} \neq \kappa_{p_4}$ of 
the derivative 
currents $j_\mathbf{p}$ and $j_{p_4}$
we are anticipating medium induced Lorentz symmetry breaking.
The relation between the 
parameters $\kappa_\mathbf{p}^2/\kappa_{p_4}^2=3$
restores Lorentz symmetry in the vacuum. 

Within the mean-field 
approximation
the regularized thermodynamic potential 
takes the following form
\be
\Omega = \Omega_\mathrm{cond}+\Omega_\mathrm{kin}^\mathrm{reg}+
\Omega_\mathrm{free}^\mathrm{reg}~,
\label{eq:ome}
\ee
\be
\Omega_\mathrm{cond} =
\frac{1}{2G_S}\left(\sigma_B^2 + \kappa_\mathbf{p}^2 \sigma_A^2
 + \kappa_{p_4}^2 \sigma_C^2 \right)-\frac{\omega^2}{2\eta_VG_S}~,
\label{eq:cond}
\ee
\be
\Omega_\mathrm{kin}^\mathrm{reg}=- N_f N_c \int\frac{d^4p}{(2\pi)^4}
\mathrm{tr}_D\log\left[\frac{S^{-1}(\tilde{p})}{S_0^{-1}(\tilde{p})}\right]~,
\label{eq:kin}
\ee
and
\bea
\Omega^\mathrm{reg}_\mathrm{free} &=&
- \frac{N_f N_c}{24\pi^2} \Big\{ 2\tilde{\mu}^3 \tilde{p}_F
- 5m^2 \tilde{\mu} \tilde{p}_F \nonumber \\
&+& 3m^4 \log\left(\frac{\tilde{p}_F+\tilde{\mu}}{m}\right) \Big\}~,
\nonumber \\
\quad \tilde{p}_F &=& \sqrt{\tilde{\mu}^2-m^2}~.
\label{eq:free}
\eea
%%%%
Here
\be
S^{-1}(\tilde{p})= -(\boldsymbol{\gamma}\cdot\mathbf{p})~ A(\tilde{p}^2)
-\gamma_4 \tilde{p}_4 C(\tilde{p}^2)+B(\tilde{p}^2)~,
\label{eq:qprop}
\ee
is the dressed quark propagator with the scalar and vector dressings
\bea
A(p^2) &=& 1+\sigma_A f(p^2),\nonumber \\
B(p^2) &=& m+\sigma_B g(p^2),\nonumber \\
C(p^2) &=& 1+\sigma_C f(p^2)
\eea
and $S_0(p)$ is the free quark propagator.
We have introduced a shorthand $\tilde{p}=(\mathbf{p},\tilde{p}_4)$
where $\tilde{p}_4 = p_4-i\tilde{\mu}$ and $\tilde{\mu}=\mu-\omega$.

The usage of the non-local approach has the important 
advantage of fitting
the form-factors $g(p^2)$ and $f(p^2)$ to lattice data
for dynamical scalar $B(p^2)$ and vector 
dressing $A(p^2)$ \cite{Parappilly:2005ei,Kamleh:2007ud,Noguera:2008cm}.
Let us also mention that the strong infrared running of the QCD
correlation functions is a core feature of the Dyson-Schwinger
approaches \cite{Fischer:2009gk,Roberts:2012sv}. 
In this work we adopt the 
parametrization from \cite{Noguera:2008cm}.

The model is solved by finding the 
extremum of Eq.~(\ref{eq:ome}) with respect 
to the mean-fields
$X=\sigma_{A},\sigma_B,\sigma_C,\omega$
\be
\frac{\partial\Omega}{\partial X} = 0~.
\ee
The EoS is obtained from evaluating the thermodynamic
potential at the extremum
\be
p = -\Omega+\Omega_0~,
\ee
where the constant $\Omega_0$ ensures 
zero pressure in the vacuum.

\subsection{\label{sec:level2} Equation of state for hadron phase \\
---Br\" uckner-Hartree-Fock theory}
Our theoretical framework for the hadron phase of matter is the 
nonrelativistic Brueckner-Hartree-Fock approach~\cite{bhf} based on the 
microscopic nucleon-nucleon~($NN$) potentials. 
The Br\"uckner-Hartree-Fock calculation is a reliable and well-controlled 
theoretical approach for the study of dense baryonic matter. 
The detailed procedure can be found in 
Refs.~\cite{Schulze:1995jx,Baldo:1998hd,Vidana:1999jm}. 
In this paper, we do not consider hyperon degrees of freedom, since they are 
superseded by the existence of quarks as we suggested~\cite{Maruyama:2007ey}. 

For the $NN$ interaction we adopt the 
so-called Bonn-B~(BOB) potential \cite{Machleidt:1987hj}.
We also use semi-phenomenological Urbana UIX nucleonic three body 
forces~(TBF)~\cite{Pudliner:1997ck}.
The nucleon mass is given as $m_n=m_p=939$ MeV.

\begin{figure}[htb]
\includegraphics[width=0.47\textwidth]{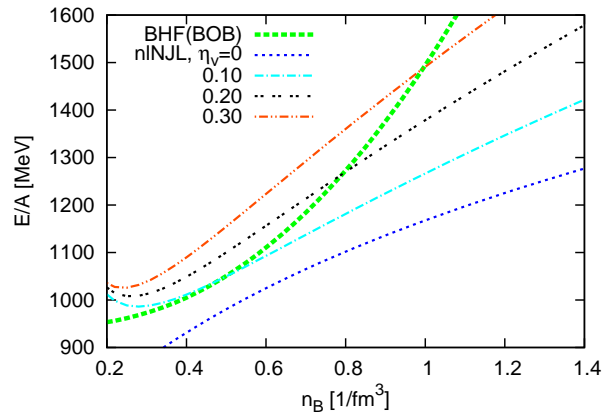}
\caption{(Color online) Energy per baryon $E/A$ in comparison 
with pure hadron(thick curve) and quark phases(thin curves) for 
different values of the vector coupling $\eta_V$ as 
indicated on the plot.}
\label{fig:eoa}
\end{figure}
On Fig.~\ref{fig:eoa} we show the resulting EoS of the pure hadron 
(thick curve with the caption ``BHF(BOB)") and the pure quark phase 
(thin curves with the caption ``nlNJL" and labels
for different values of the vector coupling $\eta_V$).
Within this model approach we find that the vector coupling of the quark phase 
must be $\eta_V \geq 0.10$, otherwise the energy per baryon number $E/A$ of 
the quark phase is lower than $E/A$ in the hadron phase.
On the other hand, very recent theoretical arguments 
\cite{Steinheimer:2014kka} point that the vector channel interaction for quark 
matter might be small.
Therefore, in this work we adopt $\eta_V=0.10$ as the lowest value and 
$\eta_V=0.20$ as the highest value.

\subsection{Hadron-quark mixed phase under the Gibbs conditions}
\label{s:mix}

To take into account the finite-size effects, 
we impose the Gibbs conditions on the mixed phase~\cite{Heiselberg:1992dx},
which require the pressure balance and the equality of the chemical potentials
between two phases besides the thermal equilibrium.
We employ the Wigner-Seitz approximation in which the whole space 
is divided into equivalent cells with given geometrical symmetry, 
specified by the dimensionality 
$d=3$ (droplet or bubble), $d=2$ (rod or tube), or $d=1$ (slab).
The structures of tube and bubble are opposite distributions of rod and 
droplet~\cite{Yasutake:2012dw}. 

The quark and hadron phases are separated in each cell with volume $V_W$ :  
a lump made of the quark phase with volume $V_Q$ is embedded
in the hadronic phase with volume $V_H$ or vice versa.
A sharp boundary is assumed between the two phases and the surface energy
is taken into account in terms of a surface-tension parameter $\sigma$.
Chiral quark model studies suggest values in the range $5-30$ MeV fm$^{-2}$ 
\cite{Palhares:2010be,Pinto:2012aq,Mintz:2012mz}, 
see also Ref.~\cite{Lugones:2013ema} where a range of $145 - 165$ MeV fm$^{-2}$
was found. 
In our calculations we use $\sigma=10$ MeV fm$^{-2}$ and  
$\sigma=40$ MeV fm$^{-2}$ and discuss the effects of its 
variation as in our previous studies with a simpler quark 
model~\cite{Maruyama:2007ey}.

We use the Thomas-Fermi approximation for the density profiles of
hadrons and quarks.
The Helmholtz free energy for each cell is then given as
\be
\begin{split}
E &= \sum_{i=n,p} \int_{V_H}\!\!\! d^3\rv\, \mathcal{E}_H[n_i(\rv)] 
 + \sum_{q=u,d} \int_{V_Q}\!\!\! d^3\rv\, \mathcal{E}_Q[n_q(\rv)] \\
 &+ E_e + E_C + E_S
\end{split}
\label{eq:e}
\ee
with $i=n,p$, $q=u,d$, 
$\mathcal{E}_H$~($\mathcal{E}_Q$) is the free energy density for 
hadron~(quark) matter, and $E_S=\sigma S$ the surface energy with $S$  being 
the hadron-quark interface area. 
$E_e$ is the free energy of the electron gas.
For simplicity, muons are not included in this paper.
The value of $E_C$ is the Coulomb interaction energy calculated by,
\be
 E_C = \frac{e^2}{2} \int_{V_W}\!\!\! d^3\rv d^3\rv'\,
 \frac{   n_{\rm ch}(\rv)  n_{\rm ch}(\rv')   }{    |\rv-\rv'|     }    \:,
\ee
where the charge density is given by 
\be
e n_{{\rm ch}} (\rv) = \sum_{i=n,p,e} Q_i n_i(\rv)
\ee
in $V_H$ and 
\be
e n_{{\rm ch}} (\rv) = \sum_{q=u,d,e} Q_q n_q(\rv)
\ee
in $V_Q$ with $Q_i$ (or $Q_q$) being the particle charge 
($Q_e=-e < 0$ for the electron).
Accordingly, the Coulomb potential $\phi(\rv)$ is defined as
\be
 \phi(\rv) = -\int_{V_W}\!\!\! d^3\rv'\, 
 \frac{   e^2 n_{\rm ch}(\rv')   }{   \left| \rv - \rv' \right|   } + \phi_0 \:,
\label{e:vcoul}
\ee
where $\phi_0$ is an arbitrary constant representing 
the gauge degree of freedom.
We fix it by stipulating the condition,  
$\phi(R_W) = 0$, 
as in Refs.~\cite{Voskresensky:2001jq,Tatsumi:2002dq,Maruyama:2005vb}.
The Poisson equation then reads
\bea
 \Delta \phi (\rv) = 4 \pi e^2 n_{{\rm ch}}(\rv) \:.
\label{e:poisson}
\eea

\begin{figure*}[htb]
\includegraphics[width=0.48\textwidth]{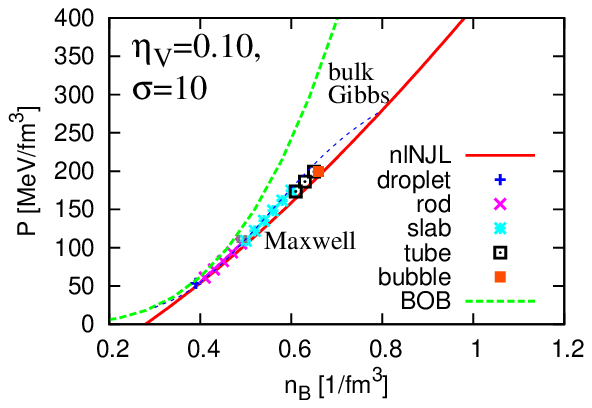}
\includegraphics[width=0.48\textwidth]{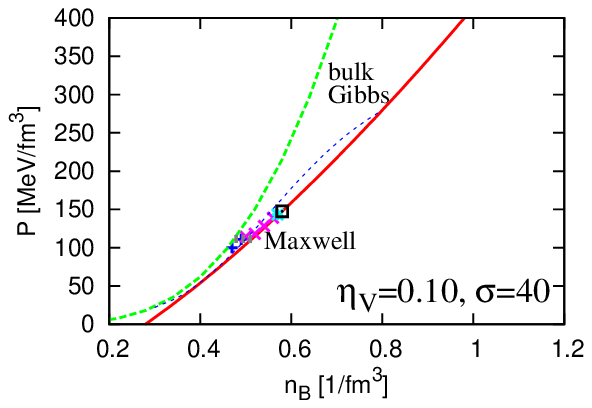}
\includegraphics[width=0.48\textwidth]{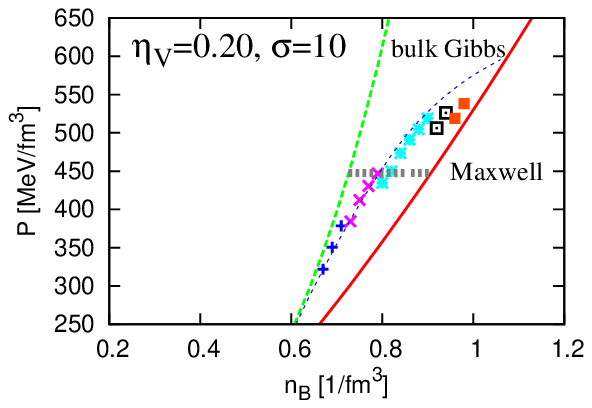}
\includegraphics[width=0.48\textwidth]{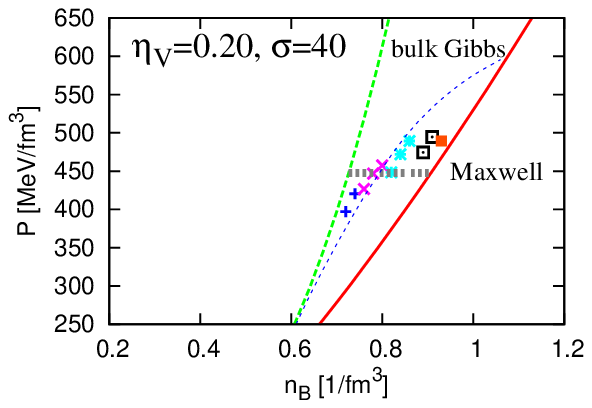}\\
\caption{
(Color online)
The EoS  in terms of pressure $p$ versus baryon density $n_B$ 
for the  pure hadronic BOB model (bold dashed line), the purely quark matter 
nlNJL model (bold solid line) and three alternatives of the mixed phase
construction: pasta phases of different structures~(symbols) in comparison 
with the Maxwell construction~(thick dotted line) and the bulk Gibbs 
construction~(thin dashed line). 
The left (right) panels show results for the same surface tension 
$\sigma = 10~(40)$ MeV  fm$^{-2}$. 
The upper (lower) panels are for the same vector coupling 
$\eta_V=0.10~(0.20)$. 
}
\label{fig:eos}
\end{figure*}

Under the Gibbs conditions, we must consider chemical equilibrium at the 
hadron-quark interface as well as inside each phase
\bea
 && \mu_u+\mu_e = \mu_d\:,  
\nonumber\\
 && \mu_p+\mu_e = \mu_n  = \mu_u + 2\mu_d \:. 
\label{e:chemeq}
\eea
For a given baryon number density 
\be
 n_B = \frac{1}{V_W}\! \left[
 \sum_{i=n,p} \int_{V_H}\!\!\!\! d^3\rv n_i(\rv)
 +\! \sum_{q=u,d} \int_{V_Q}\!\!\!\! d^3\rv \frac{n_q(\rv)}{3} \right] \:,
\ee
Eqs.~(\ref{e:poisson}--\ref{e:chemeq}), 
together with the global charge neutrality condition,
$$\int_{V_W}\!\!{d^3\rv} n_{{\rm ch}}(\rv)=0~,$$ 
obviously fulfill the Gibbs conditions.

%%%%
\subsection{Strangeness in compact stars?}

The question arises whether it is customary to generalize the approach to the 
three-flavor case before attempting a comparison of results with compact star
observables.
We argue that our present restriction to the two-flavor case in the hadronic 
as well as in the quark matter phase in this work may be considered quite 
reliable. 
Starting with increasing density in hadronic matter one should expect the onset
of hyperons to play a role for the compact star structure. 
It turns out, however, that the appearance of hyperons leads to a softening of 
neutron star matter which lowers considerably the maximum mass in 
contradiction with the observation of pulsars with masses of $\sim 2~M_\odot$
 \cite{Demorest:2010bx,Antoniadis:2013pzd}. 
This problem is known as the hyperon puzzle and its standard solution consists
in circumventing the appearance of hyperons in neutron star matter by an early
transition to quark matter \cite{Baldo:2003vx}.
For a recent discussion see \cite{Lastowiecki:2011hh,Zdunik:2012dj}, 
and references therein.
The occurrence of the strange quark flavor in quark matter, on the other hand
is shown to be a sequential process within chiral quark models
\cite{Gocke:2001ri,Blaschke:2008br,Blaschke:2010vd}. 
The reason for the sequential deconfinement in those models is that as a 
necessary condition the value of the quark chemical potential has to exceed
that of the dynamically generated quark mass of a given flavor. 
As a result, the strange quark matter phases appear at higher densities than 
the two-flavor quark matter. 
Actually, as has been demonstrated before \cite{Klahn:2006iw}, the onset of 
strangeness in cold quark matter leads to a softening of matter which in 
particular for the three-flavor color superconducting (CFL) phase entails the 
instability of hybrid star configurations beyond that threshold. 
The onset of strange quark matter thus marks the end of the stable hybrid star
configurations (maximum mass star). 
Following these arguments the structure of stable compact star configurations  
may well be devoid of strangeness in hadronic as well as in quark matter 
phases.

\section{Numerical Results}
\label{sec:results}

\subsection{Effects of the surface tension and the vector coupling
on the EoS and on the finite-size structure}

Using above relations, we study the hadron-quark mixed phase.
The four panels of Fig.~\ref{fig:eos} show the resulting pressure of the 
hadron-quark mixed phase in comparison with that of the pure hadron and quark 
phases in the relevant range of baryon density for 
$\sigma = 10~(40)$ MeV fm$^{-2}$~on the left (right) panels and for 
$\eta_V=0.10~(0.20)$~on the upper (lower) panels.
The bold dashed and solid lines indicate the pure hadron and quark phases,
respectively, while the symbols indicate the mixed phase in its various geometric 
realizations obtained by the full calculation.
The transitions between the different geometrical structures are, 
by construction, discontinuous and a more sophisticated approach 
would be required for a more realistic description of this feature.
For comparison, the hadron-quark phase transition resulting from
the Maxwell construction is shown by the thick, dotted gray line and the 
result of the bulk Gibbs construction by the thin, dashed blue line.

Compared with the case of weak surface tension ($\sigma=10$ MeV fm$^{-2}$), 
the mixed phase with strong surface tension ($\sigma=40$ MeV fm$^{-2}$)  
is restricted to a smaller density interval and the EoS gets closer to the 
one given by the Maxwell construction, even though we properly apply the 
Gibbs conditions. 
This reduction of the mixed phase region due to the charge screening 
and surface tension effects has already been demonstrated earlier
for the case when a simple bag model is used for describing quark matter
\cite{Voskresensky:2001jq,Voskresensky:2002hu,Endo:2005zt,Maruyama:2007ey}. 
%We have shown that the geometrical structure also becomes unstable due to 
%the effects of the surface tension.

As shown in Fig.~\ref{fig:eoa}, the free energy per baryon $E/A$ of the nlNJL 
model becomes large for strong $\eta_V$. 
Hence we observe that the region of the mixed phase shifts to higher densities 
as the vector interaction is increased, see also Fig.~\ref{fig:eos}. 
This behavior implies that the density region of the mixed phase becomes 
larger for stronger $\eta_V$.

\begin{figure}[tb]
\includegraphics[width=0.44\textwidth]{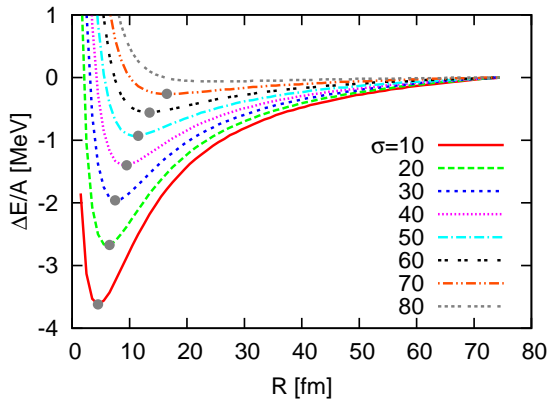}
\caption{
(Color online) The dependence of the free energy per baryon on the
droplet radius $R$ at $n_B$= 0.70 fm$^{-3}$ for different surface tensions. 
The quark volume fraction $(R/R_W)^3$ is fixed to the optimal value at 
$\sigma=10$~MeV fm$^{-2}$ for each curve. 
The free energy is normalized by its value at $R\rightarrow\infty$.
Filled circles on each curve shows the minimum energy configuration. 
}
\label{fig:stable}
\end{figure}

\begin{figure}[t]
\includegraphics[width=0.48\textwidth]{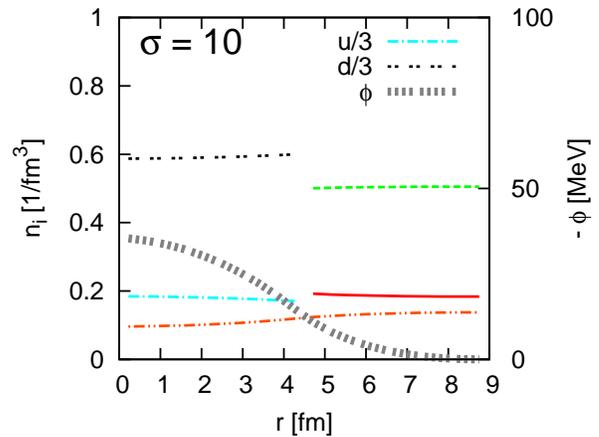}
\includegraphics[width=0.48\textwidth]{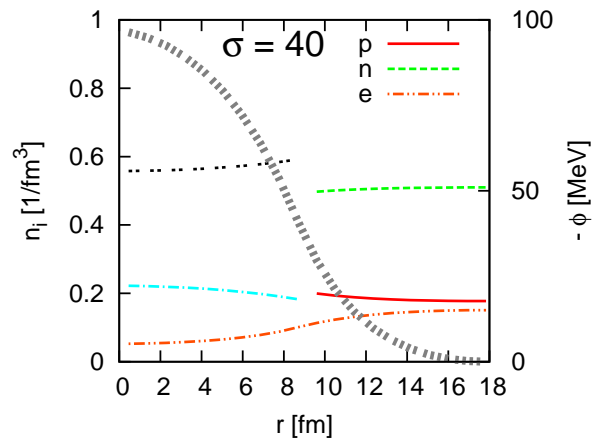}
\caption{
(Color online)
Density profiles and Coulomb potential $\phi$ for a 3D~(quark droplet) when 
$n_B=0.70$ fm$^{-3}$ for $\sigma = 10$ MeV fm$^{-2}$~(upper panel) and for
$\sigma=40$ MeV fm$^{-2}$~(lower panel). 
The cell size is $R_W= 8.98 $~fm~($18.3$~fm) with a droplet radius 
$R=  4.50 $~fm~($9.14$~fm) for the surface tension 
$\sigma=40$~MeV fm$^{-2}$ ($10$~MeV fm$^{-2}$).
}
\label{fig:prof}
\end{figure}

Fig.~\ref{fig:stable} shows the free energy per baryon of the droplet structure
for several values of surface tension at $\eta_V=0.20$. 
The quark volume fraction $(R/R_W)^3$ is fixed to exclude the trivial 
$R_W$ dependence. 
Here we use, for example, the optimal value of $(R/R_W)^3$ at 
$\sigma=10$~MeV fm$^{-2}$ for all curves. 
We normalize them by subtracting the free energy at infinite radius,  
$\Delta E/A = ( E-E(R\rightarrow\infty) )/A$,
to show the $R$ dependence clearly. 
Optimal sizes of $R$ can be evaluated from the minimum energy of $\Delta E/A$ 
obeying the variational principle.
The structure of the mixed phase is mechanically stable below 
$\sigma \sim 70$ MeV fm$^{-2}$. 
For larger values of the surface tension the minimum disappears so that the 
formation of finite-size structures is no longer favorable. 
The optimal value of the radius $R$ is shifted to larger values as 
$\sigma$ increases. 
This behavior is a signal of the mechanical instability resulting from the 
interplay between charge screening and surface tension effects.
To elucidate this point more clearly, we discuss the contribution of $E_S$ to 
$E$ in Eq.~(\ref{eq:e}). 
For a given quark volume fraction $\lambda = (R/R_W)^3$, the contribution of 
the surface  energy on $E/A$ is defined as  
\be
E_S/V_W \sim \lambda \frac{\sigma}{R},
\ee
where $V_W$ is the volume of the Wigner-Seitz cell. 
Hence, it is simply understood as  $E_S/A \sim 1/R$. 
To reduce the total energy $E$, a large $R$ is favored for a strong surface 
tension, which means that the effects of surface tension increase $R$. 
Similar results have been obtained in previous studies, although they adopted  
a simple bag model for the quark matter EoS
\cite{Voskresensky:2001jq,Voskresensky:2002hu,Endo:2005zt,Maruyama:2007ey}.

\begin{figure*}[tbh]
\includegraphics[width=0.48\textwidth]{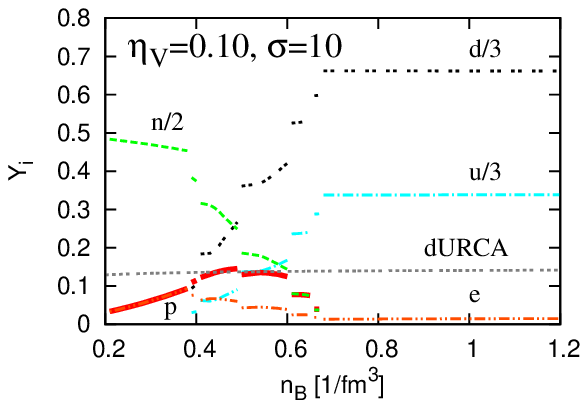}
\includegraphics[width=0.48\textwidth]{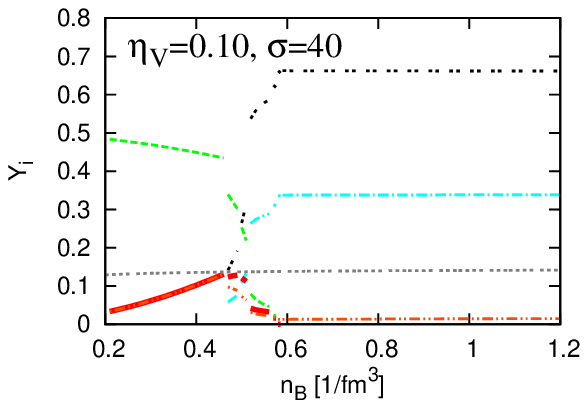} \\
\includegraphics[width=0.48\textwidth]{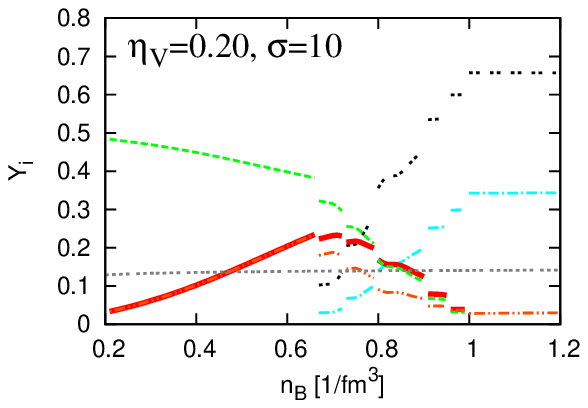}
\includegraphics[width=0.48\textwidth]{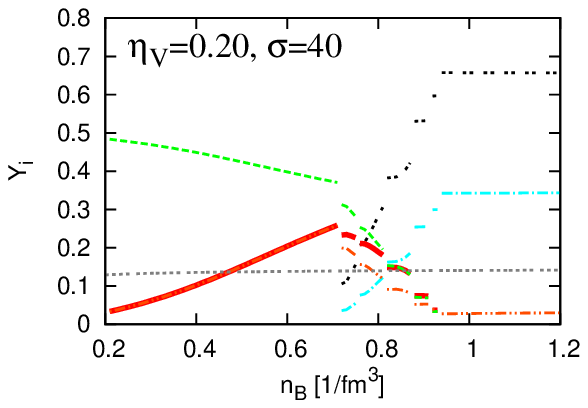} \\
\caption{
(Color online)
Left (right) panels show the particle fractions at weak (strong) surface 
tension $\sigma =10$ MeV fm$^{-2}$ ($40$ MeV fm$^{-2}$).
The upper (lower) panels have weak (strong) vector couplings 
$\eta_V$=0.10 (0.20).
}
\label{fig:Yi}
\end{figure*}

When we treat the Coulomb potential and the charge densities in a 
self-consistent manner, we can see the charge screening effect. 
It gives rise to the Debye screening mass for the Coulomb interaction and 
induces the rearrangement of charge densities.
In Fig.~\ref{fig:prof}, the density profiles within a 3D cell (quark droplet) 
is shown for $n_B= 0.70$ fm$^{-3}$ with weak (strong) surface tension, 
$\sigma=10$~MeV fm$^{-2}$~($40$~MeV fm$^{-2}$)~in the upper (lower) panel. 
We also fixed the vector interaction as $\eta_V=0.20$, the same value as in
Fig.~.\ref{fig:stable}.
The electron density is continuous in this figure. 
But all the other densities are not since a sharp boundary is assumed between 
the two phases.

Although the values of the quark volume fraction $(R/R_W)^3$ and the cell 
sizes $R_W$ are fixed in Fig.~\ref{fig:stable}, their optimal values are also 
evaluated by the variational principle as shown in Fig.~\ref{fig:prof}.
The optimal cell sizes are $R_W= 8.98$ fm for $\sigma=10$~MeV fm$^{-2}$, and 
$R_W=  18.3$ fm for $\sigma=40$~MeV fm$^{-2}$.
The optimal droplet radii are $R= 4.50$ fm for $\sigma=10$~MeV fm$^{-2}$, and 
$R= 9.14$ fm for $\sigma=40$~MeV fm$^{-2}$.
Clearly, for strong surface tension $R_W$ is larger than for weak surface 
tension.
The fraction $(R/R_W)^3$ depends only on the total baryon density. 
Hence, when the baryon density is conserved, $R_W$ grows with $R$ according 
to the increase in $\sigma$ which is large for large $\sigma$ as shown in 
Fig.~\ref{fig:stable}.  
The effects of surface tension increase $R_W$  mainly through the 
rearrangement of the charge densities. 
Since the properties of hadron matter inside the mixed phase are very 
different from those of pure hadron matter, hadron matter is positively 
charged and its partial density drops to zero in the mixed phase.

 \begin{figure*}[hbt]
\includegraphics[width=0.44\textwidth]{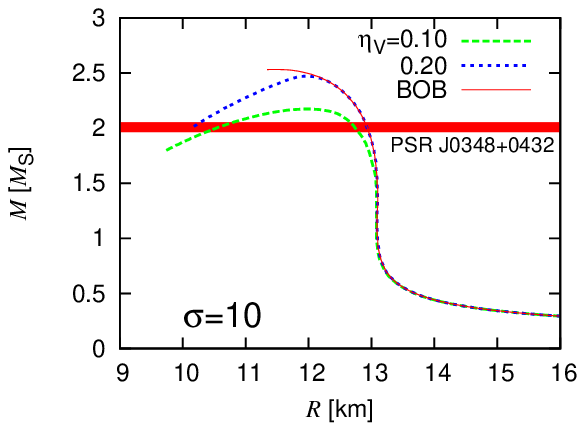}
\includegraphics[width=0.44\textwidth]{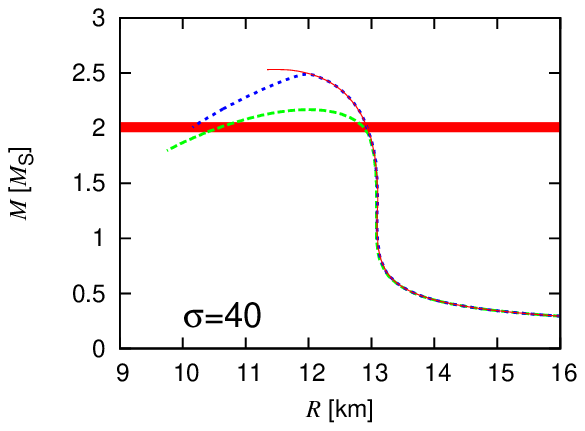} \\
\includegraphics[width=0.44\textwidth]{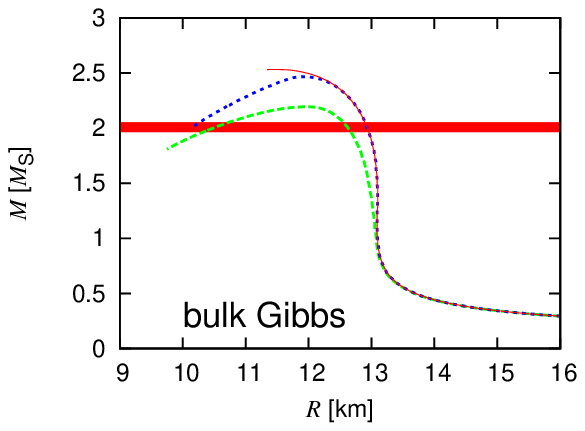}
\includegraphics[width=0.44\textwidth]{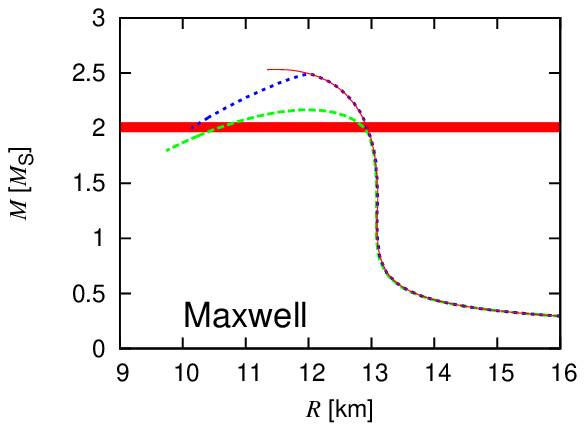} 
\caption{ (Color online)
Mass-radius relations for the choice of EoS as shown in Fig.~\ref{fig:eos}. 
The shaded area with shows the constraint of the mass measurement
$M = 2.01 \pm 0.04 M_\odot$ obtained by 
Antoniadis et.~al~\cite{Antoniadis:2013pzd} 
from observational date for the pulsar PSR J0348+0432.
}
\label{fig:mr}
\end{figure*}

\subsection{Effects on the particle fraction}

Particle fractions of quark and hadron species are shown in Fig.~\ref{fig:Yi}.
Left (right) panels show results for the weak (strong) surface tension with 
values $\sigma=10$ MeV fm$^{-2}$ ($40$ MeV fm$^{-2}$) for comparison. 
Upper (lower) panels are calculated with the vector coupling $\eta_V=0.10$ 
($0.20$).
The discontinuities in the fractions are visible on all panels, since the 
mixed phases assume fixed geometrical symmetries. %, specified by the 
%dimensionality $d=3$ (droplet or bubble), $d=2$ (rod or tube), or $d=1$ (slab) 
%in this study. 

We also show by the dotted, gray line the criterion for the onset of the dURCA 
process in $n-p-e$ matter, the threshold value  $Y_p=1/9$ for proton fraction.
For proton fractions exceeding this value the dURCA process occurs
which leads to rapid cooling of the neutron star in contradiction with 
observations, see e.~g. Refs.~\cite{Page:2000wt,Popov:2005xa,Blaschke:2006gd}.
One possible resolution to this problem is that the nuclear matter
is superseded by quark matter as the density increases.
See, e.g., the case of DBHF with Bonn-A in Ref.~\cite{Klahn:2006iw}.
In the present case, with the vector channel strength $\eta_V=0.10$
the thick, red curve on Fig.~\ref{fig:Yi} shows the proton fraction
below the dURCA value.
For this picture to actually work, the dURCA process in quark matter needs to 
be suppressed, which can be accomplished by small quark pairing gaps that do 
not significantly influence the EoS \cite{Grigorian:2004jq}.
Since at this stage our calculation of the quark phase does not take
quark pairing into account our results are to be regarded as illustrative.
If the vector coupling is stronger, $\eta_V=0.20$, the onset of
quark matter is delayed and the proton fraction exceeds the dURCA value. 

\subsection{Mass-radius and mass-central density sequences}

 \begin{figure*}[hbt]
\includegraphics[width=0.44\textwidth]{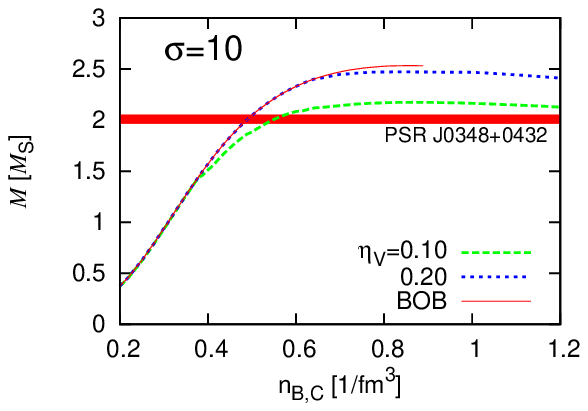} 
\includegraphics[width=0.44\textwidth]{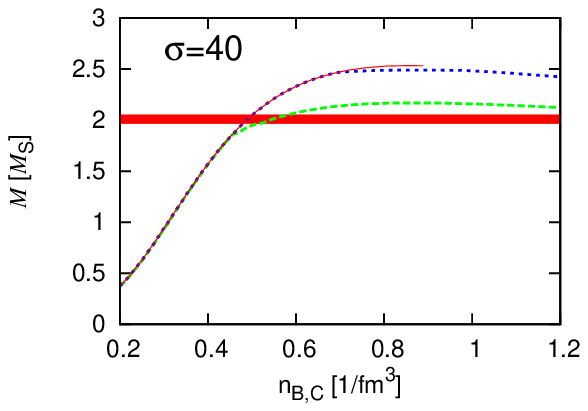} \\
\includegraphics[width=0.44\textwidth]{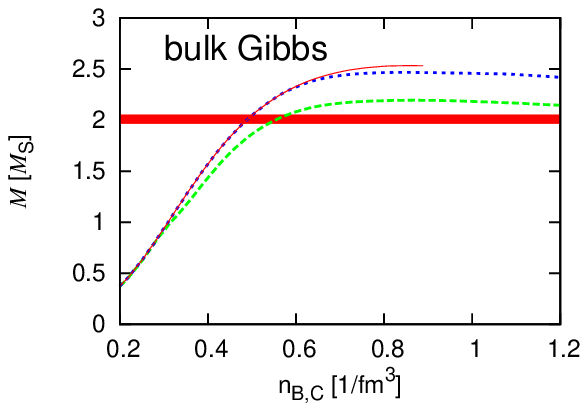} 
\includegraphics[width=0.44\textwidth]{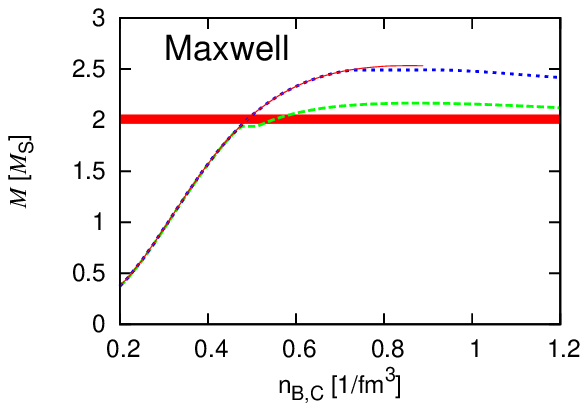} 
\caption{ (Color online)
Same as Fig.~\ref{fig:mr}, but for mass - central density sequences. 
}
\label{fig:md}
\end{figure*}

In this section we discuss some implications of our results for the EoS on the 
maximum mass of neutron stars. 
We show the mass -- radius ($M-R$) relations and the mass -- central density 
($M-n_{B,C}$) relations for isothermal hybrid stars in Fig.~\ref{fig:mr} and 
Fig.~\ref{fig:md}, respectively, obtained by solving the Tolman-Oppenheimer-Volkoff 
equations. 
Below the subnuclear density~$n < 0.1$ fm$^{-3}$, we use the BPS 
EoS~\cite{Baym:1971pw}.
In the figures \ref{fig:mr} and \ref{fig:md}, we also show
the mass range $M = 2.01 \pm 0.04 M_\odot$ obtained from observational data 
for the pulsar PSR J0348+0432 by Antoniadis et.~al~\cite{Antoniadis:2013pzd}.
All our models are clearly consistent with their result, and consequently also
with the former high-mass constraint of  $M = 1.97 \pm 0.04 M_\odot$ derived 
from observational data for PSR J1614-2230 by 
Demorest et al.~\cite{Demorest:2010bx}.

We can see that the maximum mass $M_{\rm max}$ at the weak vector coupling
$\eta_V=0.10$ is slightly smaller than that at $\eta_V=0.20$ for 
each surface tension. 
E.g., for $\sigma=10$ MeV fm$^{-2}$ we have $M_{\rm max}=2.17 M_{\odot}$ at 
$\eta_V=0.10$ and  $M_{\rm max}=2.49 M_{\odot}$ at $\eta_V=0.20$. 
This result is easily understood from the fact that strong vector interactions 
increase the stiffness of the EoS.
However, they are also responsible for the shift in the onset
of the pure quark (or the mixed) phase as understood from 
the shift in the chemical potential.
In addition, it turns out that the latter has a stronger impact on the 
stability of the star than the former.
The trend is that for higher vector couplings the appearance of quarks makes 
the star unstable.
For example, with $\eta_V=0.20$ the hybrid star branch lies on the borderline 
of stability as seen from Figs.~\ref{fig:mr} and \ref{fig:md}.

Finite-size effects have a strong influence on the 
composition of the star.
It is interesting to consider the case of $\eta_V=0.10$.
By comparing the bulk Gibbs construction from Fig.~\ref{fig:mr} with 
Fig.~\ref{fig:md} we deduce that the stable region of 
the hybrid star branch has only a mixed phase of quarks and nucleons
since the central densities for which pure quark matter phases appear 
lie on the unstable branch.
If we allow for finite-size effects, the same comparison leads to the
conclusion that the hybrid branch contains stars with a mixed
phase but also with a pure quark matter core.
In particular, the cases of $\sigma=10$ MeV fm$^{-2}$ 
and $\sigma=40$ MeV fm$^{-2}$ as well as 
the limiting Maxwell construction have pure 
quark matter in the core of the heaviest stars. 
On the other hand, we find that finite-size effects have a small
influence on the maximum mass.
For the bulk Gibbs constructions~(the extreme case of vanishing surface 
tension) the maximum mass is $2.19 M_{\odot}$ at $\eta_V=0.10$, and with the 
EoS under the Maxwell construction~(the extreme case of strong surface tension)
it is $2.17 M_{\odot}$ at same $\eta_V$.

In Table \ref{tab:mr}, we summarize the important quantities as discussed in 
this section categorized by the conditions of the phase transition and the 
strength of vector coupling. 
The conditions of the phase transition consist of the full calculations 
including the finite-size effects with the surface tension 
($\sigma =10,~40$ MeV fm$^{-2}$), the bulk Gibbs and the Maxwell constructions.
As described in Sec.~\ref{msec:eos}, we discussed as a weak vector coupling the 
value $\eta_V=0.10$, and as a strong one $\eta_V=0.20$. 
The columns ``$n_1$" and ``$n_2$" show the onset densities of the mixed and 
the pure quark phase, respectively. 
The maximum masses are shown by ``$M_{\rm max}$" in units of the solar mass 
$M_\odot$.  
The column ``$n_{dU}$" shows the onset densities of the dURCA process. 
The last columns show the size of the pure quark matter $r_2$, that of the 
mixed phase $r_1$ and the radius $R$ for four different masses of the star:
1) the mass of a typical binary radio pulsar $M_{\rm BRP}=1.4~M_\odot$, 
2) the presently largest of precisely measured masses 
$M_{\rm Antoniadis}=2.01~M_\odot$, 3) the mass at the onset of a pure quark
matter core $M_2$ and 4) the maximum mass for the given case. 
Inspecting these results we make a few observations for the hybrid EoS 
presented in this work.
At $M_{\rm BRP}$, there is no pure quark matter core in compact stars.
Even a mixed phase core does not occur, except for the extreme case of a bulk 
Gibbs construction for the weak vector coupling case, when it extends over a 
little more than half the radius. 
At the mass of the Antoniadis pulsar, $2.01~M_\odot$, for the weak vector 
coupling a quark matter core is expected in all cases (but only for the Maxwell
construction this is pure quark matter) while for the stronger vector coupling
there is none. 
When the mass is increased to that of the maximum stable configuration, then
also in the case of a structured mixed phase a pure quark matter core is 
formed which extends over almost half the star's radius, followed by a mixed 
phase layer of thickness depending on the surface tension. From zero thickness
for the Maxwell construction over $1.8$ km for the larger surface tension to
$3.9$ km for the small surface tension. For vanishing surface tension in the 
bulk Gibbs construction case the whole quark matter core is in the mixed 
phase and extends up to $9.9$ km, i.e. over more than $3/4$ of the maximum
mass star with a radius of $12.3$ km.   

\begin{table*}
\caption{\label{tab:model} 
A summary of our main results for hybrid star parameters with structures 
(pasta) in the mixed phase, categorized by the relative vector interaction 
strength $\eta_V$ and the surface tension $\sigma$.
For comparison, the extreme cases of the phase transition under the bulk Gibbs 
and the Maxwell constructions are shown which do not contain 
the finite-size effects, neither surface tension $\sigma$ nor Coulomb 
interaction. 
}
\begin{ruledtabular}
\begin{tabular}{lcccc|ccc|ccc|ccc|cccc}
$\eta_V$ & $n_1$ & $n_2$ &  $n_{dU}$ & $M_{\rm max}$/$M_\odot$ & 
\multicolumn{3}{c}{At $M=M_{\rm BRP}$}& \multicolumn{3}{c}{At $M=M_{\rm Anton}$}& \multicolumn{3}{c}{At $M=M_{2}$} & \multicolumn{3}{c}{At $M=M_{\rm max}$} \\
&  [fm$^{-3}$] &  [fm$^{-3}$]  &  [fm$^{-3}$] &   & $r_2$[km] & $r_1$[km] & $R$[km] & $r_2$ & $r_1$ & $R$ & $r_2$ & $r_1$ & $R$ & $r_2$ & $r_1$ & $R$ &\\
\hline            
\multicolumn{18}{c}{{\bf bulk Gibbs condition}} \\
0.10 & 0.30 & 0.80 & 0.41 & 2.17 
& - & 7.19 & 12.98 & - & 7.93 &12.61 & \multicolumn{3}{c|}{unstable} & - & 9.91 & 12.31 \\     
0.20 & 0.61 & 1.06  & 0.48 & 2.47  
& - & - & 13.09 & - & - & 12.93 & \multicolumn{3}{c|}{unstable} & - & 5.56 &11.92\\   
\hline
\multicolumn{18}{c}{{\bf full calculation with pasta structures, $\sigma$ = 10 MeV fm$^{-2}$}} \\
0.10 &  0.39 & 0.67 & 0.46 & 2.17 
& - & - & 13.09 & - & 7.57 &12.74 & 0 & 8.25 & 12.45 & 4.39 & 8.48 & 11.98 \\     
0.20 &  0.67 & 1.00  & 0.48 & 2.47 
& - & - &13.09 & - & - & 12.93 & \multicolumn{3}{c|}{unstable} & - & 4.25 &11.96\\ 
\hline        
\multicolumn{18}{c}{{\bf full calculation with pasta structures, $\sigma$ = 40 MeV fm$^{-2}$}} \\
0.10 &  0.47 & 0.58 & - & 2.17 
& - & - & 13.09 & -& 4.32 & 12.87 & 0 & 5.78 & 12.65 & 5.41 & 7.20 &11.98\\
0.20 &  0.70 & 0.94 & 0.48 & 2.49  
& - & - & 13.09 & - & - &12.93 &  \multicolumn{3}{c|}{unstable} & - & 3.01 & 11.96\\
\hline                             
\multicolumn{18}{c}{{\bf Maxwell construction}}\\
0.10 & 0.48 & 0.51 &  - & 2.17 
& - &- & 13.09 & 3.33 & 3.33 &12.88 & 0 & 0 &12.96 & 6.76 & 6.76 &11.84\\  
0.20 & 0.73 & 0.91 & 0.48 & 2.49  
& - & - & 13.09  & - & - &12.93 & 0 & 0 & 12.03 & 0 & 0 & 12.03\\  
\end{tabular}
\end{ruledtabular}
\label{tab:mr}
\end{table*}

\section{Conclusions and Discussion}
\label{sec:conclude}
We have studied the hadron-to-quark-matter phase transition with 
{\it finite-size effects} by imposing the Gibbs conditions on the phase 
equilibrium, and calculated the density profiles in a self-consistent manner. 
For the quark phase we used the covariant nonlocal NJL model, while the hadron 
phase was given by the BHF EoS with Bonn-B potential.

At strong surface tension, the EoS of the hadron-quark phase transition gets 
close to that given by the Maxwell construction.
This is due to the mechanical instability of the geometrical structure induced 
by the surface tension.  
The pressure of the mixed phase shows a similar behavior to that of the 
bag model
\cite{Voskresensky:2001jq,Voskresensky:2002hu,Endo:2005zt,Maruyama:2007ey}. 
It appears that this behavior of the hadron-quark phase transition is 
universal.
Since the EoS has many uncertainties, especially concerning quark matter we 
plan to study this behavior using other quark and hadron models such as 
\cite{Chen:2013tfa}.
Moreover, color superconductivity may also change our results
\cite{Pagliara:2007ph,Klahn:2013kga}.

We have found that the models used here describe compact star sequences with 
maximum masses exceeding the present constraint of $\sim 2M_\odot$ as deduced 
from observations \cite{Demorest:2010bx,Antoniadis:2013pzd}.
For the low value of the vector coupling quarks appear at low densities,
which might work in favor of suppressing the dURCA cooling channel
in nuclear matter by the early onset of quark matter.
If we conjecture that future observations would find quark matter
in neutron stars our results would indicate that vector interactions
in the quark phase are small, unlike the ones in nuclear matter.

Since the phase transition to quark matter leads to a 
softening of the EoS, this
is usually associated with the reduction in the maximum mass.
The fact that the nuclear EoS employed in 
this work gives a neutron star surpassing $2M_\odot$
therefore works in favour of obtaining a sufficiently heavy hybrid star
to exceeding $2M_\odot$ as well.
However, we should keep in mind that this 
statement depends on the relative stiffness of the nuclear
and the quark EoS at the highest densities reached in the core.
There are notable exceptions in the literature, 
see e.~g.~\cite{Baldo:2003vx,Maruyama:2007ey,Lastowiecki:2011hh,Zdunik:2012dj,Alford:2013aca} 
where one finds an opposite scenario so that 
the maximum mass of the hybrid star
is actually larger than the maximum mass of the pure nuclear star.

In the context of the microscopically founded EoS model presented in this paper with the BHF approach to the hadronic phase and the nonlocal chiral quark model for the deconfined phase, the nonstrange hybrid star scenario appears as the most conservative one.
Scenarios including strangeness in the hadronic and/or quark matter phase may require additional 
stiffening effects, that are beginning to be explored \cite{Benic:2014iaa}, 
in order to meet the $2M_\odot$ mass constraint.
We shall return to such scenarios in subsequent work.

\begin{acknowledgments}
We are grateful to H.~J.~Schulze, G.~F.~Burgio, M.~Baldo for their warm 
hospitality and fruitful discussions.
This work was supported by JSPS KAKENHI Grant Numbers 25105510, 23540325, 24105008,  
and by the Polish National Science Centre (NCN) under grant number UMO-2011/02/A/ST2/00306.
S. B. is supported by the University of Zagreb under Contract No. 202348
and by the MIAU project of the Croatian Science Foundation. D. B. acknowledges acknowledges Grant No. 1009/S/IFT/14 529
from the Polish Ministry of Science and Higher Education 530
(MNiSW).

\end{acknowledgments}

\end{document}